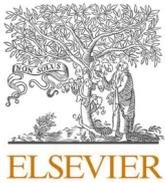
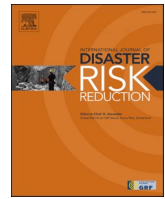

# Empirical insights for designing Information and Communication Technology for International Disaster Response

Milan Stute [a,*], Max Maass [a], Tom Schons [a], Marc-André Kaufhold [b], Christian Reuter [b], Matthias Hollick [a]

[a] *Secure Mobile Networking Lab, Technical University of Darmstadt, Germany*
[b] *Science and Technology for Peace and Security, Technical University of Darmstadt, Germany*

A B S T R A C T

Due to the increase in natural disasters in the past years, Disaster Response Organizations (DROs) are faced with the challenge of coping with more and larger operations. Currently appointed Information and Communications Technology (ICT) used for coordination and communication is sometimes outdated and does not scale, while novel technologies have the potential to greatly improve disaster response efficiency. To allow adoption of these novel technologies, ICT system designers have to take into account the particular needs of DROs and characteristics of International Disaster Response (IDR). This work attempts to bring the humanitarian and ICT communities closer together. In this work, we analyze IDR-related documents and conduct expert interviews. Using open coding, we extract empirical insights and translate the peculiarities of DRO coordination and operation into tangible ICT design requirements. This information is based on interviews with active IDR staff as well as DRO guidelines and reports. Ultimately, the goal of this paper is to serve as a reference for future ICT research endeavors to support and increase the efficiency of IDR operations.

## 1. Introduction

The number and impact of natural disasters around the world has increased in the last decades [19,39]. At the same time, man-made disasters and high-impact natural disasters such as hurricane Maria (2017) as well as the Haiti (2010) and Tohoku (2011) earthquakes created new challenges for International Disaster Response (IDR). As a result, Disaster Response Organizations (DROs) need to adapt and improve coordination, communication, and other forms of supporting technologies. Disaster Response Organizations (DROs) already aids in coordination and communication, however, current technologies such as Internet forums, satellite communications, and analogue handheld radios are rather old (yet, simple and reliable). Novel technologies have the potential to greatly improve disaster response efficiency. For example, communication systems for challenged environments based on Disruption-Tolerant Networks (DTNs) can help to collect information and communicate in the field even in the absence of cellular infrastructure [33]; cloud-based information aggregators and hubs have the potential to improve IDR coordination; and cyber-physical systems (search-and-rescue robots, drones) can support the relief efforts in the field [6,41].

To allow adoption of these novel technologies, ICT system designers and developers have to take the particular needs of DROs into account and consider the challenged environment in which they operate. We believe that this inevitably requires solid background knowledge in the field of IDR. Unfortunately, the humanitarian sector and ICT research are only loosely related fields, which leads to the problem that, on the one hand, the humanitarian sector does not know what ICT can accomplish and, on the other hand, ICT can only guess the requirements of IDR. In this work, we address the latter by building a bridge between ICT experts and the world of IDR.

In particular, we (1) provide a reference on IDR concepts and terminology; (2) define disasters and their categories including examples; (3) give an overview on DROs' roles and operations on the international and local level; and, finally, (4) translate the specifics of DRO coordination and operation into tangible and actionable insights (marked as INSIGHTS in the text).

With this, our paper enables ICT experts to design systems according to the particular needs of IDR. For example, our insights help to understand the information flow and interfaces necessary to build effective communication systems. To this end, we were in contact with 126 IDR experts and active staff (including 15 interviews) from 71 different national and international DROs who also provided pointers to official guidelines, reports, and other resources.

This article is structured as follows: In the next section, we outline the body of related work. We present our methodology in Section 3. In Section 4, we present our empirical insights for ICT design. In particular, we introduce important terminology in Section 4.1. We then explain the organizational structure of DROs at the example of the United Nations (UN) in Section 4.2. We detail how different DROs coordinate on a global level in Section 4.3, and how a typical local IDR operation is coordinated and carried out in Section 4.4. Finally, we conclude in Section 5.

## 2. Related work

The field literature refers to ICT for disaster response as crisis informatics, although the terms disaster, crisis, and emergency are often used interchangeably. Due to the growing number of crisis situations occurring across the world [20,25], the use of crisis communication and management via technology has gained in importance and been increasingly researched [46,47]. There are several challenges and obstacles in sharing and coordinating information during multi-agency disaster response [8].

Research efforts have focused on describing the specific characteristics of emergencies and the resulting challenges and requirements for ICT support, for example, derived via case studies and interviews [11,37]. The conference for Information Systems for Crisis Response and Management (ISCRAM) was the first scientific venue for ICT-based crisis communication. Founded in 2004 by a group of scientists from various related fields, its aim was to address the issue of ICT support for effective [69] disaster management. One major work presented at the first meeting was a 'dynamic emergency response management information system' for stakeholders from different areas [56]. As part of this work, crisis characteristics and their management have been addressed, as well as behaviors and needs of the affected parties.

Other scholars have highlighted long-term consequences and high risks of damage (i.e., lack of resources or casualties), contributing to even more complex crisis situations due to interconnectedness of infrastructures and systems [11,37]. At the same time, some works have shed light on issues of limited time for planning, necessary communication before responding, and the public demand for timely and reliable information [11,37,56]. Furthermore, because crisis situations are relatively rare, sudden and unpredictable, there is a high degree of uncertainty [11,37]. While this inherent uncertainty always restricts perfection of preparation efforts, it is still decisive whether staff has been trained using ICT before an emergency [11,37,56]. At the same time, various parties are involved in and affected by crisis response and decision-making, so that a conflict of interests and a need for negotiation between the stakeholders may arise [9,11,37,56]. Research has also been conducted on individual stakeholders' roles in an emergency, being dynamic and hard to predict whereas sharing information without barriers is said to be essential [56].

Communication via ICT has been used to analyze and address these diverse stakeholder requirements [9]. Diverse tools and channels like social media, mobile apps, online maps and smart watches have been utilized to facilitate information and communication between authorities, emergency responders and citizens [10,15,38,48,65]. Based on their premises concerning emergencies, Turoff et al. [56] derived requirements and principles for the design of ICT which should serve as an orientation when developing for crisis support. They focus on a "group communication system" which should include "metaphors, roles, notifications, context visibility, and hypertext" [56]. Such a system relies on a hierarchical, interconnected and retrievable data structure. Time-, location- and recipient-specific notifications and information are required, which should be up-to-date, context-based and respondable.

Moreover, ICT systems for crisis support have to provide flat and barrier-free communication, appropriate role and task assignment, and support of team-building.

Some authors highlight the need for resilient ICT [2]. Crisis information sources [31,36], platform tools [14] and, in general, a trend towards mobile applications [54] and cloud computing [57] are omnipresent, which may also lead to technology related risks [17].

Despite a broad body of research conducted on the nature of crises and crisis management support via ICT, we found that there is a dearth of studies defining concrete characteristics and requirements particularly for ICT designers. Most of the work refers to specific events, considers only a particular group of stakeholders, or develops and tests prototypes or frameworks. However, this encompasses only a part of ICT potential for crisis management.

## 3. Methodology

The aim of the empirical study was to understand IDR and eventually derive our insights for ICT support. As shown in Fig. 1, our approach is structured in two different qualitative methods that were used to obtain results and to gather information:

(1) document analysis: DRO policies, guidelines, strategies and field books partly collected from online disaster response platforms, and
(2) interviews: individual on-site and remote interviews with disaster relief experts.

The main reason to use document analysis and interviews as two different qualitative methods was to reconstruct the practices. Therefore the usage of the open coding of the grounded theory [13] was helpful to analyze the material and to uncover interesting phenomena. Furthermore, it is important to emphasize that the data collected has a distinct EU-focus and is particularly relate to Germany and outlaying countries and thus tend to cover a small geographical area.

### 3.1. Sampling process

During the process, on of the authors was responsible to contact a total of 298 IDR experts and received feedback from 126 of them within 71 organizations (42% response rate). In particular, we asked about (1) operational and final reports as well as debriefings; (2) best practices, also first-hand experiences; (3) strategies and procedures for handling disasters; (4) information about GPS positioning data from cars and employees in crisis and disaster areas; and (5) search, positioning, and supply strategies.

From the responders, we received additional information about their humanitarian activities, past mission reports, and pointers to other experts and contacts better suited for replying to our inquiry. In addition, we received access to exclusive online platforms and forums such as the Virtual On-Site Operations Coordination Center (VOSOCC) of the UN Office for the Coordination of Humanitarian Affairs (OCHA), reliefweb. int, and humanitarian response.info. They contain field handbooks and guidelines for in-field operations which are not publicly accessible.

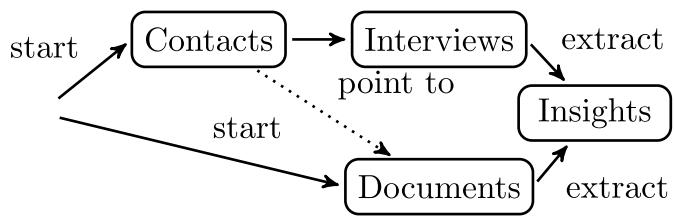

**Fig. 1.** Our process for information gathering, interviews, and insight extraction.

Furthermore, we were able to conduct 15 on-site and Skype interviews with IDR experts and former disaster relief workers. Overall, we received feedback from individuals associated with the following national and international organizations: Caritas (AT), Berufsfeuerwehr Bochum (DE), Johanniter (DE), Red Cross (DE), Technisches Hilfswerk (DE), emergency.lu (LU), Humanitarian Intervention Team (LU), Ministry of Foreign Affairs (LU), Concern Worldwide, Plan International, ZOA International, Emergency Telecommunication Cluster (ETC), Global Disaster Alert and Coordination System (GDACS), Office for the Coordination of Humanitarian Affairs (OCHA), United Nations Organizations (UNO), United Nations Office for Disaster Risk Reduction (UNISDR), World Food Program (WFP), and many others who wished to remain anonymous.

### 3.2. Data collection of the interviews

Before being able to produce this work, we had to choose a different approach for researching the required background information. Initially when we started our research, we scanned any website, guideline, policy, or strategic document from any DRO we could find. However, we realized that research solely based on this kind of information would probably still rely on a large amount of assumptions taken by researchers not yet completely familiar with the topic. Therefore, we chose to take our research one step further and tried to personally get in touch with experts from the field of IDR. The overall intention is to gradually reduce the amount of assumptions used when creating DTN simulation models for post-disaster areas. We think that by getting in touch and cooperating with experts having first-hand experience and a broad knowledge about this specific area, we moved in the right direction. Some of the experts from the field of IDR not only provided us with insights into their daily work but were willing to help us get access to exclusive online platforms used by disaster response experts from around the globe. This enabled us to get an even more "real-time" insight into IDR since we had the chance to read live discussions of IDR experts debating on how to proceed in real disaster scenarios happening at this very moment. We thus had the opportunity to read comments and thoughts of experienced disaster response managers sharing their ideas and experiences on disaster preparedness and disaster response planning with others. Furthermore, we were able to download internal documents such as field handbooks and guidelines for in-field operations (which were also not publicly accessible). The document analysis was especially useful to gather extensive background information on IDR and DRO, including the UNO's efforts on an international level as well as the cooperation of governmental and non-governmental organizations. However, we decided to go one step further and to conduct one-on-one (one-site) interviews. Out of the received 126 responses, we were able to conduct 15 interviews (see Table 1). We carried out semi-structured expert interviews that allowed us to gain a deeper understanding of the topic. Most of the interviews were conducted via Skype or telephone. However, some of them were also conducted on site. The length of the interviews varied from 15 min to 2 h. All interviewees were experts working in the area of IDR or DRO. The main objective was to gain additional insights and further clarifications on their daily work and past experiences with disaster relief activities. Therefore, the interviewees were asked four leading questions about (1) their emergency experiences and their respective roles, (2) the daily operation in these emergencies, (3) locations of team sites (such as base camps), and (4) recurring (technical) problems during emergencies.

These leading questions were designed to identify problems with existing ICT systems as well as define the geographic conditions potentially affecting radio communications. In addition, knowledge about these geographic conditions combined with user mobility information can help to derive disaster-specific mobility models [53]. Conducting semi-structured interviews, we followed these leading questions and gained a deeper understanding of the topic, for instance, allowing answer-dependent courses of conversation to emerge.

### 3.3. Data analysis

As mentioned before, the data analysis was based on the inductive approach found in the grounded theory approach [13]. This facilitates the development of a broader theory. To be able to use this methodology, the material was coded openly and the statements of the agents were divided into text modules and later into categories. In particular, through the possibility of open coding, a large amount of data from the analysis of the documents as well as the interviews could be processed. The open coding process is characterized by the division of the data into segments. Afterwards, it is scrutinized for commonalities that reflect categories or themes. Generally, the overreaching goal of open coding is to reduce the amount of data to a small set of themes.

Based on this grounded theory-based analysis, the researchers jointly analyzed the data in four steps to construct a new theory that is grounded in that data. In our study, the collected data is gained both through expert interviews as well as document analysis. The analysis of the collected data started based on the question of what challenges of designing ICT systems have to be met in the context of the particular needs of IDR. All three researchers performed an intensive analysis of the documents and literature provided as well as the conducted interviews with disaster relief experts. They identified different repeating themes, which they, afterwards, coded with keywords and phrases. Through the relationship identification of the concepts, the different insights identified during the coding process (see Table 5) have been categorized in five main categories: Terminologies, Disaster Types, Disaster Organizations, International Coordination and Local Cooperation and Operation. These categories, as well as the links found between them, are used as the basis for the development of a new theory presented in Section 4. Nevertheless, there is no written code book, but only a derivation of the categories from the collected data and the literature analysis.

We chose this systematic methodology to discover insights about the interviewees' daily work practices and past experiences with disaster relief activities through the analysis of data. The knowledge previously acquired in the literature study was used to heighten theoretical sensitivity [13]. A part of the grounded theory approach is theoretical sampling, which means that the selection of the studied units is led by the conceptual structure or theory that emerges during the analysis. Essentially, the analysis process is characterized by the development of categories and their subsequent grouping into meta categories, which can be seen in the next section.

## 4. Results

Based on our interviews and collected material, we attempt to extract insights for ICT design. We first discuss important terminology. Then, we analyze disaster response organizations as well as international coordination. Finally, we look at the local operation.

**Table 1**
List of interviewed organizations out of 71 organizations in total.

| No | Interview partner |
|---|---|
| 1 | SOS Children's Villages International |
| 2 | Plan International |
| 3 | Ministère des Affaires étrangères et européennes |
| 4 | INSARAG (International Search and Rescue Advisory Group) |
| 5 | CARITAS Austria |
| 6 | Emergency Directorate in Concern Worldwide |
| 7 | ZOA |
| 8 | University of Stuttgart, Department for Visualization and Interactive Systems (VIS) |
| 9 | Fire Service Bochum |
| 10 | German Red Cross |
| 11 | World Food Programme Rome |
| 12 | Luxembourg Humanitarian Intervention Team |
| 13 | Luxembourg Humanitarian Intervention Team |
| 14 | Caritas Germany, Disaster Relief Coordination Unit |
| 15 | St. John Accident Assistance Germany |

*4.1. Use of terminology*

This section explains basic terms and the way they are used in practice.

*4.2. Disaster*

In the field of IDR, we find four seemingly similar but distinct terms: *hazards*, *disasters*, *crisis*, and *emergencies*. A *hazard* can be defined as the probability of occurrence, within a specific period of time in a given area, of a potentially damaging natural phenomenon [60] or, more recently, as a dangerous phenomenon, substance, human activity or condition that may cause loss of life, injury or other health impacts, property damage, loss of livelihoods and services, social and economic disruption, or environmental damage [61]. A *disaster*, on the other hand, can be considered as the manifestation of a hazard, i.e., a serious disruption of the functioning of a community or a society involving widespread human, material, economic or environmental losses and impacts, which exceeds the ability of the affected community or society to cope using its own resources [61,63]. In this work, we are only concerned with *disasters*. However, we do not exclude the scientific debate and practical implementations regarding *crises* and *emergencies* due to their interchangeable use and interconnectedness. While all four concepts refer to an event of disruption, emergencies and disasters are sometimes distinguished from one another because emergencies can be managed following standard procedures while disasters (and crises) often require extraordinary measures [1,12].

The UN defines a disaster cycle which is composed of *response*, *recovery*, *mitigation*, and *preparedness* [43]. In this work, we are primarily concerned with response meaning the provision of assistance or intervention during or immediately after a disaster to meet the life preservation and basic subsistence needs of those people affected [43]. At the same time, we consider the phase of preparedness (i.e., activities prior to disasters), directly linked with the shape of disaster response [43].

Many Non-Governmental Organizations (NGOs) have conceptualized more, and more specific, terms. These include, for example, the phases of search and rescue, emergency relief, early recovery, medium to long-term recovery, and community development [58]. This difference is due to the fact that many NGOs provide long-term support and remain active in the country long time after the initial disaster had struck, thus reflecting the context-dependency of categories.

*4.3. Categories and types*

Literature [45, 68] as well as the UN and major NGOs differentiate between two main categories of disasters: *man-made* (also called *anthropogenic* or *technological*) disasters and *natural* disasters. Man-made disasters are caused by humans and occur mainly in, or close to, human settlements. This can include environmental degradation, pollution, transportation and industrial (e.g., nuclear) accidents, as well as (armed) conflicts. Natural disasters relate to disasters not caused by human actions but by natural physical phenomena which can be of geophysical, hydrological, climatological, meteorological, or of biological nature [7, 68]. Examples include storms,[1] floods, and earthquakes. While it is of course comprehensible to distinguish between human-made and natural disasters, one has to keep in mind that the two types and their respective origins may not be strictly isolated from each other. Table 2 illustrates this categorization. The main categories are further divided into *foreseeable* and *sudden-onset*, which separates events that can be predicted with high accuracy and from those that cannot be predicted or only a few hours or days in advance, respectively. Examples for foreseeable natural disasters usually include droughts and other reoccurring seasonal weather phenomenon, while earthquakes and tsunamis can be considered sudden-onset events. Note that a clear distinction between foreseeable and sudden-onset events is not always possible because technical monitoring and alert systems can be unreliable, for example, when predicting volcanic activity.

---
[1] The terms *hurricane*, *cyclone*, and *typhoon* all refer to the same type of storm but are used depending on the storm's location: in the Atlantic and Northeast Pacific, in the Northwest Pacific, and in the South Pacific and Indian Ocean, respectively.

The *EM-DAT* [19] and *NatCatSERVICE* [39] databases use the same categories and types and contain valuable information about past disasters and their impact. The former is maintained by the Center for Research on the Epidemiology of Disasters (CRED) and holds more than 17000 entries dating back to 1900. The latter is maintained by Munich RE (a German reinsurance company) and holds more than 26000 entries, dating back to 79 AD. Table 3 is an excerpt of these databases. It lists recent large-scale natural disasters and shows that the most severe ones are most frequently caused by earthquakes and storms.

While natural disasters are most devastating, our interview partners mentioned that especially man-made disasters were hard to predict due to sudden occurrence and their especially erratic character, leading to difficulties in adapting to the situation.

**Insight 1.** *Uniqueness of disaster.* While there exist disaster classifications, many factors including scale and local conditions make every instance unique. IDR and, in extension, ICT needs to prepare for a variety of different events and local conditions, but also the ability to quickly adapt to unforeseen situations. This translates to several system requirements such as flexibility and ease of deployment, which we pick up in the following insights.

*4.4. Disaster response organizations*

In this section, we explain key elements of DROs, with a particular focus on the UN. We provide a general background on the nature of these organizations and explain their strategies, plans, and setup. Furthermore, we highlight the coordination mechanisms, responsibilities, and respective duties of selected organizations. The primary goal of disaster response is to save lives. In the often chaotic environment of a disaster, response efforts require clear structures and assignment of responsibilities to assert coordinated and organized operation between the different relief units. Fig. 2 provides an overview of the different actors involved in IDR and their relations. It is intended to provide visual guidance throughout the entire section.

*4.5. The United Nations*

**Table 2**
Disaster categories and examples.

| Category | | Example |
| --- | --- | --- |
| Natural | Geophysical | volcanic activity, earthquake, landslide, tsunami |
| | Hydrological | flood, avalanche |
| | Climatological | extreme temperature, drought |
| | Meteorological. | cyclone, storm surge, wildfire |
| | Biological | disease epidemic, insect and animal plague |
| Man-made | | armed conflict, industrial accident, pollution |

We explain the role of the UN in disaster relief activities, and highlight some specific departments within the UN and their particular tasks. Furthermore, we provide the reader with background information on terms and strategies required for the understanding of the later sections.

UN agencies have pre-existing development-focused relationships with member states of the UN, and thus, provide sector-specific support

**Table 3**

Natural disasters since 2010 (excerpt). Note that a study on hurricane Maria estimates a corrected death toll of more than 3000 [21].

| Disaster | Year | Death toll | Area (km²) |
| --- | --- | --- | --- |
| Hurricane Maria | 2017 | 64 | 10 063 |
| Nepal earthquake | 2015 | 9 000 | 3 610 |
| Cyclone Pam | 2015 | 24 | 12 190 |
| Ludian earthquake | 2014 | 617 | 1 487 |
| Typhoon Haiyan | 2013 | 6 300 | 71 503 |
| Christchurch earthquake | 2011 | 185 | 1 426 |
| East Africa drought | 2011 | 260 000 | 2 346 466 |
| Tropical storm Washi | 2011 | 1 292 | 104 530 |
| Tohoku earthquake | 2011 | 15 894 | 83 955 |
| Haiti earthquake | 2010 | 316 000 | 27 750 |

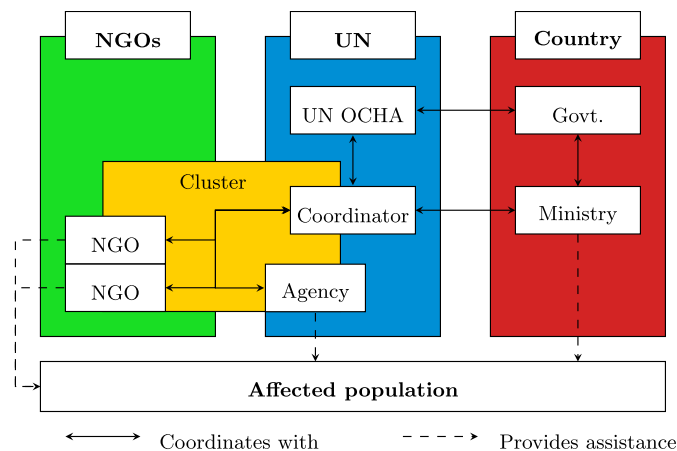

**Fig. 2.** Overview of entities involved in IDR (simplified and non-exhaustive).

and expertise before, during, and after a disaster. The most senior UN official in a country is usually designated as the Resident Coordinator (United Nations Regional Coordinator (UN RC)) and acts as the primary focal point for a government's engagement with the UN system. Within the UN itself, several sub-agencies are in charge of dealing with the effects of international disasters.

As coordination is key in chaotic situations, the UN has implemented a *cluster* system with the 2005 Humanitarian Reform [27]. The cluster system defines eleven different key activities and the respective groups therein.[2] Humanitarian organizations (UN and non-UN such as NGOs) form clusters working in the same main sectors of humanitarian action to increase transparency and accountability, enhance predictability, have more effective advocacy, and align their efforts (joint strategic and operational planning) [24]. Each cluster has a single point of contact: the cluster *coordinator* who reports to the cluster lead agency. Clusters are created when clear humanitarian needs exist or national authorities demand to do so.

**Insight 2.** *Interfaces.* IDR involves many independent and interdependent groups. In a disaster, it is imperative to immediately find the right contact. Thus, IDR employs precise interfaces between the different stakeholders. For example, the UN RC acts as the interface between a government and the UN. Supporting ICT systems should acknowledge these interfaces and facilitate use by all parties involved.

Following the dominant UN approach of the disaster programme cycle and its four phases while focusing on disaster preparedness as well as response, we introduce the main actors (summarized in Table 4) in charge of the two respective areas of disaster management within the UN.

---

[2] The eleven key activities of the cluster system are: logistics, nutrition, emergency shelter, camp management & coordination, health, protection, food security, emergency telecommunication, early recovery, education, and sanitation water & hygiene [24].

The United Nations Office for Disaster Risk Reduction (UNISDR) prepares for and reduces risks of disasters. The UNISDR campaigns for more disaster resilience, particularly in poor countries, and advocates climate change mitigation and sustainable development [64].

The campaigns for more d develops procedures for search-and-rescue operations. INSARAG attempts to render emergency preparedness and response activities more effectively, improve the cooperation efficiency amongst international USAR teams, promote activities designed to improve search-and-rescue preparedness in disaster-prone countries, and develop USAR procedures, guidelines, and best practices [26].

The Office for the Coordination of Humanitarian Affairs (OCHA) globally coordinates IDR activities. At field level, OCHA ensures that the humanitarian system functions efficiently. OCHAs's other functions include inter-cluster coordination, developing policies, managing information, and organizing field support and humanitarian financing. In order to do this, OCHA has regional and country offices worldwide and can deploy additional staff at short notice. Especially during emergencies, when local capacity is overwhelmed, OCHA can deploy their staff within hours such that specialists can help to coordinate the disaster relief activities. The United Nations Disaster Assessment and Coordination (UNDAC) dispatches professional first responder teams. In the disaster area, UNDAC teams assess, coordinate, and manage information. Moreover, they set up and manage the On-Site Operations Coordination Center (OSOCC) which acts as an information exchange and coordination hub (see Section 4.4). Deployed UNDAC teams are self-sufficient for up to one week, use pre-defined methods and procedures, and are usually entirely composed of experienced emergency managers.

**Insight 3.** *Self-sufficiency.* Deployed teams bring with them all (or at least most of) the equipment they need to carry out their operation so that dependence on local resources (Insight 5)) is minimized. This is necessary because supplies might be delayed, damage assessment has not been carried out and, therefore, it is uncertain which resources are available on site. ICT systems should therefore be powered by self-sustaining energy sources such as photovoltaic panels in combination with batteries. Backup power system should be provisioned.

Similar to OCHA and UNDAC, World Food Programme (WFP) teams are ready to be deployed upon government request. Especially in the early days of an emergency and in sudden-onset disasters, WFP teams assess and quantify the exact amount of food assistance required, as well as for how many beneficiaries and for how long such help needs to be sustained [66]. Furthermore, WFP spots humanitarian corridors for food delivery, and their logistic teams organize an immediate shipment. To transport food to disaster areas WFP uses ships, planes, helicopters, trucks, sometimes even donkeys and yaks, depending on the geographic conditions. In addition, our interview partners mentioned that, depending on the country, infrastructure hubs such as hospitals, airport, or city hall provide best communication means.

**Table 4**
UN departments and their tasks.

| Organization | Description |
| --- | --- |
| UNISDR | prepares for and reduces risks of disasters. Its main activity is to serve as the focal point in the United Nations system for the coordination of disaster reduction and to ensure synergies among the disaster reduction activities of the United Nations system and regional organizations and activities [59]. |
| INSARAG | develops procedures for search-and-rescue operations. They are a network of countries and organizations, dedicated to USAR and operational field coordination. |
| OCHA | coordinates humanitarian actors and ensures coherent response to emergencies [42]. They are mandated to coordinate humanitarian actions between the UN, DROs, and governments. |
| UNDAC | is designed to help the UN and governments of disaster-affected countries during the first phase of a sudden-onset emergency. They coordinate incoming international relief at national level or at the site of the emergency and can deploy teams worldwide within 12–48 h [62]. |
| WFP | aims at saving and protecting lives in emergencies, supporting food security and nutrition, and reduce undernutrition worldwide. They delivers food to more than 80 million people in 80 countries around the world. |
| ETC | provides communication services in humanitarian emergencies [16]. |

**Insight 4. *Flexibility*.** IDR needs to adapt to heterogeneous local conditions, such as geography, setting (rural or urban), or political situation in the country. Instead of "one size fits all" solutions, they need multiple different solutions to be flexible. Therefore, ICT systems need to be tailored to the scenario.

**Insight 5. *Local resources*.** If possible, locally available resources will be used to "get stuff done." These resources are often the only ones available in the immediate aftermath and are often already well-adjusted to the local conditions (climate, terrain, etc.). In addition, local infrastructure will be exploited if present. Supporting ICT should therefore be able to exploit this infrastructure, e.g., by connecting to a functional cellular network instead of only relying on ad hoc communications.

Given that ICT systems should be applicable in a variety of scenarios (Insight 4) and that system should make use of diverse local resources (Insight 5), designers face the risk of over-engineering, i.e., building a system that is too complex to be used effectively. Therefore, we believe that system should be kept as simple as possible.

**Insight 6. *Simplicity*.** Time is critical, so simple "no fuss" solutions are required. On the one hand, this means that user interfaces should be minimalistic while remaining functional. On the other hand, simplicity implies that communication systems should be ready-to-use, requiring self-configuring, self-healing, and self-optimizing networks.

The ETC provides ICT services in the field. In the early days post-disaster timely and effective information and communication technology greatly improves humanitarian response and coordination. The coordinate is part of ETC and can be dispatched within hours after notification in order to set up basic communication needs for the WFP in a disaster area within 48 h [67]. FITTEST consists of IT, electrical, and radio experts specially trained to work and operate in demanding and hostile conditions when setting up basic communication means.

Unfortunately, even disaster preparedness cannot guarantee reliable communication. For example, the 2015 Nepal earthquake had been foreseen so that preparation took place two weeks in advance. Still, the communication network went down 24 h after the earthquake occurred because the backup diesel generators of the cell towers in the mountains ran out of fuel.

**Insight 7. *Resilient communication*.** Communication is key for an effective disaster response. If the local communication infrastructure is destroyed, dedicated IDR teams will set up a temporary solution which can take up to 48 h [18,29]. This backup infrastructure is for disaster response staff only and does not "connect" the affected population. Thus, resilient, infrastructure-less, and zero-interaction ICT deployments, e. g., based on disruption-tolerant networking, can be beneficial.

### 4.6. Non-governmental organizations

The term Non-Governmental Organization (NGO) is commonly used when referring to a non-profit organization that is independent from national and international governmental organizations. Despite many NGOs can be found in the humanitarian sector, the term NGO is not exclusively coined to a humanitarian activity (for example, there exist NGOs fighting to stop environmental pollution). In this work, we focus on NGOs delivering humanitarian aid, emergency assistance, and disaster relief. Humanitarian aid is comprised of provisioning of materials, logistics, and (medical) support to individuals in need. In the context of natural disasters, the primary purposes of humanitarian aid are to save lives, alleviate suffering, and maintain human dignity.

There are no official sources for the number of humanitarian aid workers around the world, but they are estimated [4] to be around 210 000 in the year 2008, where roughly 50% are NGOs, 25% belong to the International Federation of Red Cross and Red Crescent Societies (IFRC), and 25% account to the UN.

**Insight 8. *Scalability*.** The large IDR community requires scalable (technical) solutions. This is especially true when individuals from different DROs meet the first time in unknown terrain and immediately must begin to cooperate. Scalability is also important to support crowd-sourcing in ICT applications which has become more common to solve certain IDR-related tasks in a distributed manner such as [23]. Also, scalability implies that ICT needs to be interoperable, e. g., so that local and emergency systems can complement each other.

### 4.7. International coordination

International disaster response uses various Internet-based tools and technologies to coordinate the various involved groups and organizations world-wide. In this section, we first give an overview of two key platforms (GDACS and VOSOCC) and, then, explain one key element for any disaster response mission: reconnaissance.

### 4.8. Online platforms

The Global Disaster Alert and Coordination System (GDACS) is jointly operated by the European Union (EU) and OCHA and is available at http://www.gdacs.org. The web page provides an continuously updated overview on all global alerts. The portal also provides in-depth information, such as reports, satellite images, and situations briefings. GDACS' alert system automatically broadcasts alert messages for high-profile events to its subscribers.

As part of GDACS, the VOSOCC is a website for worldwide coordination of disaster relief missions, available at https://vosocc.unocha.org (access only upon registration). The VOSOCC acts as an information hub with subgroups for specific disasters and provides an interface between the local disaster relief staff and the global IDR community.

**Insight 9. *Centralization*.** IDR coordination heavily relies on centralized IT-backed information hubs: portals and forums provide the basis for current IDR endeavors.

### 4.9. Reconnaissance

When starting an operation, emergency response teams always require up-to-date maps of the disaster area. Before making any further plans, the integrity of vital infrastructure such as roads, bridges, hospitals, and airports, has to be checked. Safe spots (or large free areas) for constructing base and refugee camps need to be determined. Furthermore, it is important to assess the damage and determine the most severely hit areas. For all of the above, satellite images are extremely useful [5].

In case of major disasters, the *International Charter on Space and Major Disasters* signed by 15 space agencies can be activated. The charter provides humanitarian organizations with free-of-charge high-resolution (up to 30 cm per pixel) satellite imagery [28].

Acknowledging the importance of map data [32,52], the NGO Humanitarian OpenStreetMap Team [23] supports local first responders by enhancing the OSM map database using satellite images and augmenting it with detailed IDR context information such as infrastructure damages. The OSM project provides a solid foundation for humanitarian efforts as it provides an open interface (Insight 2), supports crowd-sourcing efforts (Insight 8), and eventually allows local first responders to download map data for printing or off-line digital use (Insight 4 and 7).

Correct information is particularly important during the first 24 h after a disaster since this period was described as "really chaotic" by our interviewees.

**Insight 10. *Timeliness, integrity, and accuracy*.** Information is key to plan a mission and minimize uncertainties (for example, road conditions when a storm has just devastated the area). Therefore, accurate and up-to-date information is required. Especially in hostile environments, information integrity must be ensured.

*4.10. Local coordination and operation*

In this section, we are concerned with in-the-field structures and operations. We introduce key physical facilities (OSOCC and RDC) that are deployed in the field by the first arriving IDR teams. We also describe typical operations by the example of USAR. Fig. 3 depicts the relationship between local facilities and operations [44].

*4.11. On-site centers*

Developed by OCHA, the On-Site Operations Coordination Center (OSOCC) is a physical facility that was originally designed to assist affected countries in coordinating international search-and-rescue efforts following an earthquake. Due to its success, it is now the standard tool used within sudden-onset disasters of any kind. The OSOCC is established as soon as possible after a disaster occurs. The first international Urban Search and Rescue Team (USRT) or UNDAC team arriving on site usually sets up the foundations for the OSOCC [44]. The three main objectives of an OSOCC are to be a link between international responders and the government of the affected country, to provide a system for coordinating and facilitating the activities of international relief efforts at the disaster site, and to provide a platform for cooperation, coordination, and information management. The OSOCC is the first point any incoming DRO should visit after its arrival on the disaster site, as they can register there and then receive general information about the situation on site, the operations of national and international responders, and make logistical arrangements. The OSOCC guidelines [44] are a very valuable resource for further information about the OSOCC operations and its tasks during a mission. Furthermore, a VOSOCC (see Section 4.3.1) is established.

**Insight 11.** *Regular synchronization.* In the field, a central coordination hub (OSOCC) is deployed to exchange information and synchronize the different actors. This is also the place for communication with the outside world (telephone or Internet access). The actors usually visit the OSOCC at least daily, for example, for a briefing in the morning and a debriefing in the evening. Personal ICT systems and sensors such as smartphones can exploit this regular synchronization for uploading collected data to a central system for aggregation.

As a part of the OSOCC, the Reception/Departure Center (RDC) is located at the main arrival point of international DRO teams as well as relief goods. Its main purpose is to facilitate and coordinate arrivals (immigration and customs) and deployment of resources (humanitarian aid, equipment). In many cases, the RDC is located at the closest airport and, thus, possibly at a different geographic location than the OSOCC. In particular, the RDC pre-registers teams, provides an initial on-site briefing, gives directions to the OSOCC, and finally passes processed information to the OSOCC. Further details on the RDC can be found in Ref. [44].

**Insight 12.** *Well-known processes.* DROs define processes for different activities such as arrival and departure. ICT solutions can exploit this fact to streamline and facilitate certain procedures such as briefings and information aggregation.

**Insight 13.** *Automated reporting.* At the end of each day and each mission, operation reports have to be written and passed up in the organizational hierarchy. In practice, these manually-written reports are often incomplete and lack potentially important detail because time is scarce. Mobile technical solutions could facilitate and partly automate the reporting process by exploiting sensors (photos, GPS) and communication capabilities (WiFi, Bluetooth) of current smartphones. Further, natural language processing could be employed to post-process reports, e. g., in order to extract relevant data.

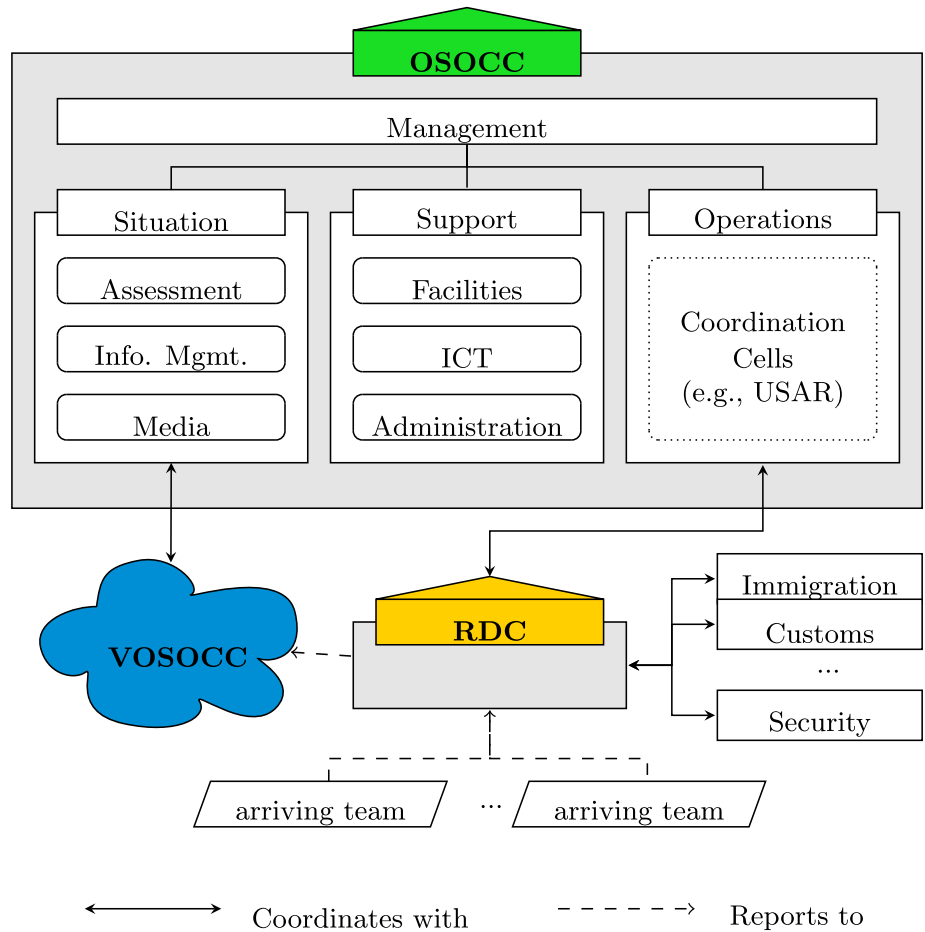

**Fig. 3.** Interplay of the OSOCC, RDC, and VOSOCC.

*4.12. Search and rescue*

Since chances to find trapped victims alive significantly decrease after a few days, speed, precision, and preparedness are of vital importance for any Urban Search and Rescue (USAR) operation. Such missions are often key to an early success in many disaster relief activities. The ultimate goal of a USAR operation is to rescue the largest number of people in the shortest amount of time, while minimizing risks for the rescuers. USRTs consist of trained experts who provide rescue and medical assistance in emergencies. International USRTs are composed of expert personnel and, depending on the requirements, specialized equipment and search dogs. They are operational within 24–48 h, whereas some are usually even faster due to permanent stand-by capacities. INSARAG specifies and certifies capacities and capabilities of international USRTs [26]. Light USRTs can assist with surface USAR, while medium USRTs can also conduct technical USAR operations in structural-collapse incidents and are adequately staffed for 24-h operations at one site for up to seven consecutive days. In addition, heavy USRTs can conduct difficult and complex technical search-and-rescue operations and search for trapped people, using canines and technical systems [43].

**Insight 14.** *Cyber-physical systems* such as unmanned ground and aerial vehicles (UGVs and UAVs) can support dangerous and problematic USAR operations avoiding human exposure to unnecessary risks such as collapsing buildings or gas leaks.

**5. Discussion and conclusion**

In this section, we summarize our results, discuss the practical implications by proposing designs for three support systems, and list the limitations of this work.

*5.1. Results*

In this work, we have introduced basic concepts of IDR. We have presented the highly heterogeneous environment in which IDR operates (Insight 1). Thorough examination of respective documents, the conduction of 15 interviews as well as granted access to guidelines for field operations allow for a better picture of strategies, procedures, and problems IDR actors encounter. Using this, we derive operation principles of IDR such as employment of clear interfaces (I. 2) and established processes (I. 12), as well as the duality of self-sufficient operation (I. 3) and use of local resources if they are available (I. 5). Information flow follows a centralized approach (I. 9) and synchronization happens regularly (I. 11). We have also extracted general system requirements such as flexibility of deployment (I. 4), ease of use (I. 6), scalability (I. 8), and the importance of timely and accurate information (I. 10) as advice for ICT system designers. Based on this, we point to three areas of research which could greatly enhance current IDR operations: decentralized and resilient means of communications, for example, based on DTNs (I. 7); USAR-supporting cyber-physical systems (I. 14); and simplifying reporting by automating parts of the process (I. 13). While some insights mentioned in this work have been mentioned in previous works, we make distinct observations that are especially relevant for and can be directly transferred to ICT system design, with the purpose of closing the respective research gap. We summarize them in Table 5.

*5.2. Implications*

In this work, we have derived principles for designing ICT systems as well as identified three concrete types of support *systems* that would benefit IDR. Those systems were mentioned during our interviews by first responders who (1) had to deal with unreliable communication, (2) spent lots of time on writing daily reports, and (3) see benefit in cyber-physical systems.

In the following, we sketch support system designs (Insights 7, 13 and 14)) that follow the operation principles (Insights 1 to 3, 5, 9, 11 and 12) and meet the requirements (Insights 4, 6, 8 and 10) of IDR.

**Resilient Communications** During disasters, communication infrastructure is often unavailable due to physical damage or overload by its users [34]. No network accesses hinders IDR operations such as search-and-rescue missions, which require up-to-date map data [3]. For coordination, point-to-point radios can be used, but their range is typically limited to a single hop and only supports voice. In addition, satellite radios are expensive to operate and, therefore, device density is typically low in the field. Self-organizing networks such as DTNs [33] where devices forward messages for one another can be deployed anywhere without any dependencies on existing infrastructure (I. 3). DTNs can be run on smartphones [35] which has the added benefit that staff can use devices they are already familiar with (I. 6). However, their range may be extended using drones [6] or—if available—even connect to cellular networks (I. 5). Technically, the challenge of deploying such networks is to accommodate a potentially large number of devices and traffic (I. 8) depending on the scale of the disaster (I. 1). Fortunately, related works [50] have shown that this can be accomplished by first and foremost designing protocols with reduced control overhead that is required to maintain a functioning network.

**Automated Reporting** IDR staff typically writes operation reports at the end of each day or mission. These report are accumulated at a central location, e. g., the OSOCC (I. 9) and passed up in the hierarchy to synchronize IDR efforts and facilitate planning (I. 11). Today, report writing is a manual, time-consuming, and therefore, often erroneous process. An automated reporting system could provide much more accurate and detailed information (I. 10) while only require minimal interactions (I. 6) from its users. In particular, we envision a system that would leverage staff equipment (smartphones) to collect sensor data such as images, location traces, and phone logs to automatically create template entries with pre-filled meta data (e. g., time and location), so IDR staff only has to add comments for that particular event. In addition, such as system should merge sensor data from different devices to generate aggregate reports so that, e. g., photos captured by any USRT member are included in a single report. Similar to Refs. [51], we suggest that such a system should draw on the extensive research body of human-computer interaction (HCI) to engage the users and design accessible interfaces.

**Cyber-physical Systems** Drones can provide assistance in IDR operations in different ways. Manual controlled drones [30] today can already provide up-to-date imagery during reconnaissance (I. 12). Video information has in the past been shown to be valuable to firefighters to assess the situation [40]. Drones can extend the imagery coverage as they can easily reach areas that are inaccessible by ground vehicles. In addition, platooning enables groups of drones to carry out more complex tasks such as providing a temporary airborne link between two DTN partitions [6] to achieve direct communication (I. 10). To train staff in using such new systems, we believe that gamification approaches can be applied [55].

**Table 5**
Summary of our insights.

| Type | No. | Name |
|---|---|---|
| Operation principles | 1 | Uniqueness of disaster |
| | 2 | Interfaces |
| | 3 | Self-sufficiency |
| | 5 | Local resources |
| | 9 | Centralization |
| | 11 | Regular synchronization |
| | 12 | Well-known processes |
| System requirements | 4 | Flexibility |
| | 6 | Simplicity |
| | 8 | Scalability |
| | 10 | Timeliness, integrity, and accuracy |
| Support systems | 7 | Resilient communications |
| | 13 | Automated reporting |
| | 14 | Cyber-physical systems |

*5.3. Limitations and future work*

While we derive design principles and propose ideas for systems that would support IDR efforts, we have neither built nor evaluated them. In addition and due to the complexity of the subject, we have focused on few but important organizations and operations involved in IDR. Yet, our work sets out to be a starting point for IDR laymen such as ICT experts and strengthen interdisciplinary research in the fields of IDR and computer science. It must also be taken into account that the paper showed the results of ICT researchers interviewing DRR people. However, there is definitely a need for further research to point out solutions on how to integrate the two worlds. In general, collaboration is necessary in the design process [22]. Tailorable solutions are needed, that allow the adaptation of the system based on the users' in-situ requirements [49]. In particular, we need to conduct realistic field trials to assess the practicality and performance of new systems and to receive tangible feedback on their usability [35].

**Declaration of competing interest**

None.

**Acknowledgments**

The authors would like to express their sincere gratitude towards all cooperating individuals from the various national and international organizations. This work has been co-funded by the LOEWE initiative within the emergenCITY center, and by the DFG as part of project C.1 within the RTG 2050 "Privacy and Trust for Mobile Users."